# EFFECTS OF SUBSTRATE ANISOTROPY AND EDGE DIFFUSION ON SUBMONOLAYER GROWTH DURING MOLECULAR BEAM EPITAXY: A KINETIC MONTE CARLO STUDY

J. Devkota
*Central Department of Physics, Tribhuvan University, Kirtipur, Nepal*

and

S.P. Shrestha[1]
*Department of Physics, Patan Multiple Campus, Tribhuvan University, Patan Dhoka, Nepal*
*and*
*The Abdus Salam International Centre for Theoretical Physics, Trieste, Italy.*

**Abstract**

We have performed Kinetic Monte Carlo simulation work to study the effect of diffusion anisotropy, bonding anisotropy and edge diffusion on island formation at different temperatures during the sub-monolayer film growth in Molecular Beam Epitaxy. We use simple cubic solid on solid model and event based Bortz, Kalos and Labowitch (BKL) algorithm on the Kinetic Monte Carlo method to simulate the physical phenomena. We have found that the island morphology and growth exponent are found to be influenced by substrate anisotropy as well as edge diffusion, however they do not play a significant role in island elongation. The growth exponent and island size distribution are observed to be influenced by substrate anisotropy but are negligibly influenced by edge diffusion. We have found fractal islands when edge diffusion is excluded and compact islands when edge diffusion is included.

[1] Corresponding author: shankarpds@gmail.com

## 1. Introduction

Recently there has been considerable interest in the studies of sub-monolayer epitaxy on metal and semiconductor surface due to their immense potential for technological applications [1-10]. Electronic devices, coatings, displays, sensors and numerous other technologies all depend on the quality of the deposited thin films. Thin film structures with smooth interfaces are of fundamental importance in the field of modern technologies related to electronics, optics and magnetism. The morphology of the films grown on the surface is highly influenced by a number of physical processes such as nucleation of deposits, surface diffusion of adatoms, formation of a superlattice at monolayer coverage etc. The choice of an appropriate substrate and deposition conditions is very essential to achieve the desired film quality. To control the quality of thin films grown by MBE, it is important to understand the role of individual microscopic processes and how they influence surface morphology. Monte Carlo, as well as Kinetic Monte Carlo (KMC) computer simulations had achieved considerable success in modeling homoepitaxial growth of metals as well as semiconductors. Using the solid-on-solid model, simulation of growth on a simple cubic lattice structure had gained considerable success in the study of the effect of various physical processes on the surface morphology of a growing film. Several groups have studied the evolution of growth of island morphologies as well as scaling of islands during submonolayer epitaxy using Monte Carlo simulation [11-16]. Using various models, the effect of anisotropy of diffusion [17-20] and binding strength [16, 20-22] between adatoms and island on island shape were studied as well**.** The influence of diffusion anisotropy and binding strength between adatoms and island border on island aspect ratio have been studied using temperature, surface diffusion barrier and adatom binding energy as parameters [22]. In this work, we present the role of diffusion anisotropy and edge diffusion on island morphology, number density and size distribution using a model for irreversible aggregation.

## 2. Growth Model and KMC Simulations

A Kinetic Monte Carlo simulation method is a probabilistic simulation method that deals with kinetic processes. Each configurational change corresponds to the real events and each event can happen with some probability that depends on their rate. Our KMC simulation model basically resembles the solid on solid model on square lattice, as used by Heyn [22]. In our simulation atoms are randomly deposited on a site of the 300x300 square lattices with the flux F on the substrate at definite temperature. The deposited atoms are allowed to hop to adjacent sites on the substrate with a rate given by

$$D_j = D_o \exp(-E_j/K_B T) \qquad (1)$$

where j = X, Y corresponds to the diffusion direction; $D_o$ is the common prefactor of order $10^{13}$ $sec^{-1}$ and $E_j$ is the barrier potential for the adatom diffusion along the respective direction. The energy barrier $E_j$ is a sum of two contributions, a site-independent surface term $E_S$ and a term given by the number of lateral nearest neighbors, $nE_N$, where $E_N$ is the in-plane bond energy. Thus,

$$E_j = E_{s,j} + nE_N \tag{2}$$

where, n = 0, 1, 2, 3, 4 is the number of occupied lateral nearest neighbors at the site. The isotropic form of energy is modified to incorporate the anisotropy of the surface with two non equivalent directions X and Y. This means the adatom can diffuse along X and Y direction with different probability. With these considerations, the energy barrier for the surface diffusion along the X-direction and Y-direction are given by

$$E_X = E_{S,X} + n_X E_{N,X} + n_Y E_{N,Y} \tag{3}$$

$$E_Y = E_{S,Y} + n_X E_{N,X} + n_Y E_{N,Y} \tag{4}$$

where $E_{S,X}$, $E_{S,Y}$ are the energy barriers for monomer diffusion in the respective direction $E_{N,X}$, $E_{N,Y}$ are bonding energies and the numbers $n_x$, $n_Y$ = 0, 1, 2 represent the number of bonds of nearest neighbors in X and Y direction respectively.

When adatom encounters another adatom as its nearest neighbor, both atoms are frozen and stable two atom island nucleates on the surface. Similarly, when a diffusing adatom encounters an existing island of size S, it sticks on the island yielding an island of size S+1. The dimers and larger islands are assumed not to move as rigid bodies. The dissociation rate of island is taken as $D_{diss,j}$ = Doexp(-$E_{diss,j}/K_BT$) along the j = X,Y direction, where $E_{diss,j}$ is the dissociation barrier along the respective direction, i.e. $D_{diss}$,j = $E_s$ + $E_{N,j}$. Since, in our model the islands formed are rigid and stable, no dissociation of atoms is allowed. This is incorporated by taking the high value of $D_{diss}$,j. Adatom on an island edge with only one bond is allowed to diffuse within the island perimeter. The hopping rate for edge diffusion is taken as De = $D_o$exp(-$E_{ed}/K_BT$), where $E_{ed}$ is the potential barrier for the edge diffusion. In our model, the rate of edge diffusion has been taken relative to the rate of monomer diffusion on the surface rather than taking an independent value of $E_{ed}$. The transition barriers for the various atomistic processes used are similar to the barriers used by Heyn [22]. It should be noted that the absolute values of temperatures are not important in the simulation since temperature scales with energy parameter. Simulation is carried out until the desired coverage is achieved and the quantities of interest are calculated. The growth temperature has been varied over a large range of 560K to 1250K in various series of simulations. The effect of substrate temperature and activation energy for terrace diffusion, edge diffusion and dissociation on island morphology, aspect ratio, and island growth exponent have been studied during our work. Due to the computer power constraint we have taken every point in the plots as the average of fifteen different ensembles only.

## 3. Results and Discussion

### Island Morphology

In order to study the influence of different processes, we have monitored the island morphology, aspect ratio, island number density and size distribution by varying different deposition parameters. The influence of anisotropic terrace diffusion is studied using a model for irreversible island growth. This is done by setting the neighbor related energy terms ($E_{N,X} = E_{N,Y} = 10$ eV) to such a high value that detachment is suppressed. Effect of edge diffusion is excluded by setting very high value of $E_{edge}$ (10 eV). The cross-channel diffusion energy barrier $E_{s,y}= 1.3$ eV is taken to be fixed and the degree of anisotropy is varied by setting the different values of in-channel substrate diffusion energy $E_{s,x}=1.3$ eV, 1 eV, and 0.8 eV. Island morphology, number density and size distribution for each energy parameter were monitored at different substrate temperatures.

Fig.1(a-l) depicts the island morphologies for three different values of in-channel diffusion barriers $E_{sx}=1.3$ eV, 1 eV and 0.8 eV for four different substrate temperatures T=650K, 715K, 835K and 1000K, respectively. The horizontal panel from left to right depicts the variation of the island morphologies with in-channel diffusion barrier for four different fixed values of the temperature (first row is for T=650K, second row is for T=715K, third row is for T=835K and fourth row is for T = 1000K). Similarly, the vertical panel from top to bottom depicts the variation of the island morphologies with the substrate temperature (T=650K, 715K, 835K and 1000K) for three different fixed values of the in-channel diffusion barrier (first column is $E_{sx} = 1.3$ eV, second column is for $E_{sx} = 1$ eV and third column is for $E_{sx} = 0.8$ eV). The study of variation of island morphology with the variation in in-channel diffusion barrier at temperature T= 650K [Fig.1 (a-c)] shows that at low diffusion barrier, the values of the island number density is low but the size of island is large. A similar trend was observed in the island morphologies for other higher temperatures. Thus, under identical condition and with the same deposition parameter, the island number density and island size depends on the type of substrate used. A similar effect was observed in different types of substrates with $E_{sx} = 1.3$ eV, 1 eV and 0.8 eV which is depicted in Fig.1.

The variation of island morphology with temperature for different fixed value of $E_{sx}$ is depicted in the vertical panels in Fig.1 (a,d,g and j). From the figure, it is observed that island number density decreases with the increase in the temperature where as the size of island is observed to increase. The island morphologies for other values of diffusion barrier also follow the similar trend as seen along the vertical panel at different energy barriers.

The observed increase in island size for substrate with higher degree of anisotropy can be explained as follows. As compared to isotropic substrate, with energy barrier $E_{sx} = E_{sy}=1.3$ eV, in anisotropic substrate with energy barrier $E_{sy} = 1.3$ and $E_{sx} = 0.8$ the diffusion barrier along X direction is low. Therefore, the adatoms make a larger number of jumps before another adatoms gets deposited. Due to this nucleation probability is decreased and growth probability is increased. Therefore, island size is increased and its number density is decreased. The result for effect of temperature can also be explained in a similar way. When the substrate temperature is high, the adatoms diffuse for longer distances within the lattice and hence they encounter frequently with the existing islands and aggregate forming a larger sized island and smaller number density.

### Island Density Exponent

Fig.2 depicts the variation of island number density with D/F ratio for isotropic [▪] and anisotropic [•] substrate. Here $D=D_o\exp(-E_{s,x}/K_BT)$ is the in-channel diffusion barrier and F is the flux. As expected, the island density was observed to decrease with the increase in D/F ratio or substrate temperature. The decrease of island density appeared continuously indicating that in the studied temperature range no change of critical nucleus size occurs. The variation of N with D/F ratio shows power law behavior. The growth exponent $\chi$ was found to be $\chi = 0.30$ and $\chi = 0.27$ for the isotropic ($E_{sx} = E_{sy} = 1.3$ eV) and anisotropic ($E_{sx} = 1$eV, $E_{sy} = 1.3$ eV) substrate, respectively. In our model, islands with size greater than or equal to 2 are not allowed to diffuse nor dissociate. Therefore, the critical nucleus size is i=1. The power law behavior of N enables us to test whether or not the scaling relation given by equation

$$N \sim \left(\frac{D}{F}\right)^{-\chi} \exp\left[\beta E_{di} /(i+2)\right] \qquad (5)$$

holds or not. Therefore, the value of growth exponent given by $\chi=i/(i+2)$ for i = 1 is expected to be $\chi=1/3$ [6, 12] for the isotropic diffusion case, and between 1/3 and ¼ for the anisotropic diffusion case [6,19]. The calculated exponents from simulation are nearly comparable to the expected values mentioned above. It was observed that the value of the growth exponent $\chi$ for isotropic substrate ($E_{sx}= E_{sy}= 1.3$ eV) is about 10% less than the theoretical value predicted by the rate equation analysis. The discrepancy is possibly due to the statistical errors in the simulation. Liu et al. [15] had obtained the value of the exponent $\chi= 0.30$ which is also 10% smaller than that of the rate equation value even at higher temperature. In their work, two atoms on the same site are assumed to form dimer, whereas¸ in the present model, two atoms in the neighbouring sites are assumed to form a dimer.

### Island Size Distribution and Scaling Behavior

Fig. 3 depicts the island size distribution $N_s$ for different values of substrate temperature T = 650K, 715K, 770K, 835K, 910K, and 1000K for the isotropic substrate with $E_{sx}= E_{sy}= 1.3$ eV. From the figure it is seen that the island size distribution is drastically influenced by the variation of the substrate temperature. It is observed that as the value of the substrate temperature increases the position of the peak of the island size distribution curve shifts towards higher island size while the peak height of the curve decreases. The width of the distribution curve is observed to broaden with the increase in the temperature.

Fig. 4 depicts the variation of island size distribution $N_s$ at different values of the substrate temperature for anisotropic substrate ($E_{sx}$= 1 eV, $E_{sy}$= 1.3 eV). The variation of $N_S$ with temperature is observed to be similar to the isotropic case. Comparing the distribution curve at the different temperature, it is observed that the distribution curve for the higher temperature is much more scattered as compared to the curve at the low temperature case. The reason for this is that at higher temperature the size of the island is bigger but the number density is much lower. Due to this, the distribution curve appears to be much more scattered.

Comparing Fig. 3 and Fig. 4 it is seen that the peak of the curve drastically decreases and shifts towards a higher size for anisotropic substrate at the same temperature. Also, the width of the peak is observed be wide. This is more clearly shown in figure 5 in which the variation of island size distribution $N_s$ for the substrate with three different in – channel diffusion energy barriers $E_{sx}$= 0.8 eV, 1 eV and 1.3 eV at fixed temperature T = 715K is plotted. Similarly, the average island sized number density is observed to change drastically with change is anisotropy of the substrate. This is because lowering the value of $E_{sx}$ to make substrate anisotropic provides more probability for an adatom to aggregate in islands. The distribution curve is also more scattered in anisotropic case than in isotropic case for the same temperature. The reason for this is that on the substrate with lower energy barrier, the probability of the encountering of two atoms increase. Due to this, the size of the island is bigger and the number density is much lower. Therefore, the distribution curve appears to be much more scattered.

Fig. 6 depict the plots of scaling function $f(s/S_{av})$ defined by the equation $N_s \approx \theta S^{-2} f(s/S)$ for the different values of the temperatures (i.e. T = 715K, 835K, 1000K) for $E_{sx} = E_{sy}$ = 1.3 eV, respectively. From the figure it is seen that, the island size distribution $N_s$ for the different temperatures plotted against scaled variable $s/S_{av}$ collapse in a single master curve showing the scaling of the island size distribution. Comparing the scaling function curve for isotropic substrate (Fig. 6) with anisotropic surface (not shown in figure), it is observed that the curve is more scattered for anisotropic surface, which is due to the similar reason explained above.

### Effect of Edge Diffusion

We have also studied the effect of edge diffusion on island morphology. The rate for an adatom to diffuse along the edge within the island was taken with respect to the rate of terrace diffusion rather than taking an independent value. In the present study, we have considered only the isotropic terrace diffusion case $E_{sx} = E_{sy}$ = 1.3 eV. Variation on island shape and growth exponent with the variation in rate of edge diffusion has been studied taking three ratios $D_e/D$ = 0.001, 0.01 and 0.1 for fixed flux 0.1 ML/s and fixed coverage 0.075 ML in the temperature range T = 600 K to T = 1000 K. Here $D_e$ is the rate of edge diffusion and D is the rate of terrace diffusion. All others phenomena such as dimer diffusion, trimer diffusion, dissociation from island edge etc. are not allowed.

**Effects on the Island Morphology**

Fig. 7 (a, b, c) depicts the island morphology at three different temperatures T = 715 K, 835 K and 1000 K for the model without edge diffusion ($D_e/D$ = 0) and Fig. 7 (d, e, f) depict the island morphologies for the same temperatures with inclusion of edge diffusion ($D_e/D$ = 0.01). Here Fig. 7 (a, b, c) is the same as Fig. 1 (a, b, c) and repeated here for comparative study. The variation of island number density and the island size increment with temperature for a model with inclusion of edge diffusion is found to be very similar to the results without inclusion of edge diffusion which is evident from the figure. Comparing the island morphologies generated with inclusion of edge diffusion, with the morphologies generated under identical growth parameter without inclusion of edge diffusion, we find that in the edge diffusion excluded case, the island shape is observed to be of dendrite nature whereas in the edge diffusion included case, the island shape is observed to be of compact nature. Thus, the effect of edge diffusion is observed to change the island shape drastically from dendrite to compact shape. From the observation of these morphologies, it is found that at higher temperature the effect of edge diffusion becomes more significant. At the higher temperature the edge diffusion becomes more effective so the islands are more compact. This is clearly seen in Fig. 7.

**Effect on the Island Growth Exponent**

Fig. 8 depicts the variation of island number density (N) as a function of substrate temperature for three different values of the ratio $D_e/D$= 0.001, 0.01 and 0.1. The island number density shows the power law behavior. From the figure, the value of growth exponent $\chi$ for $D_e/D$ = 0.001, 0.01 and 0.1 is found to be 0.30, 0.31 and 0.31, respectively, which is similar to that of the value in case of simulation without edge diffusion. The result shows that the exponent doesn't change significantly with the inclusion of edge diffusion although it changes the island morphology drastically. These results are expected because in the edge diffusion process only the edge atoms diffuse within the island. Thus the island number density remains nearly the same and hence the exponent does not change. A small change in exponent is possible. This is because at high temperature the atoms at the edge are mobile along the edge. During its motion it is probable that it may come in contact with neighbour islands which results in merging two islands into one. This leads to the decrease in number density at higher temperature. Due to this the exponent may change slightly, which is observed.

The value of the growth exponent with and without edge diffusion is observed to be nearly the same, which indicates that at the studied temperature range the edge diffusion is not operative to change growth exponent. However, it has been reported [14] that the exponent may change drastically if the rate of edge diffusion is very high using a slightly different model where edge diffusion has been allowed in small clusters like dimers and trimers. In their work, on one hand they have allowed corner rounding and in the other they have used a model in which dimer and trimer are not frozen, i.e. edge diffusion was allowed in small clusters also. Because of the constraints in computer power, we have not taken in to account these values.

### 4. Summary and Conclusion

We have presented the Kinetic Monte Carlo results of the study of the effect of edge diffusion and substrate anisotropy in terrace diffusion on the island morphology and island growth exponent for the various simulation parameters. The main findings of our works are summarized below. It is found that the anisotropy in diffusion barriers does not produce effectively elongated islands. The island growth exponent is found to be 0.30 for isotropic substrate and 0.26 for anisotropic substrate. The effect of low edge diffusion does not bring any appreciable change in island growth exponent ($\chi$) while it leads the islands to the change from fractal to the compact shape. The effect of edge diffusion is observed to be more pronounced at higher temperature.

**Acknowledgments:** One of the authors, Shankar Prasad Shrestha, is thankful to the Condensed Matter and Statistical Physics Section of the Abdus Salam International Centre for Theoretical Physics (ICTP), Trieste, Italy, for support at ICTP where part of this work is completed.

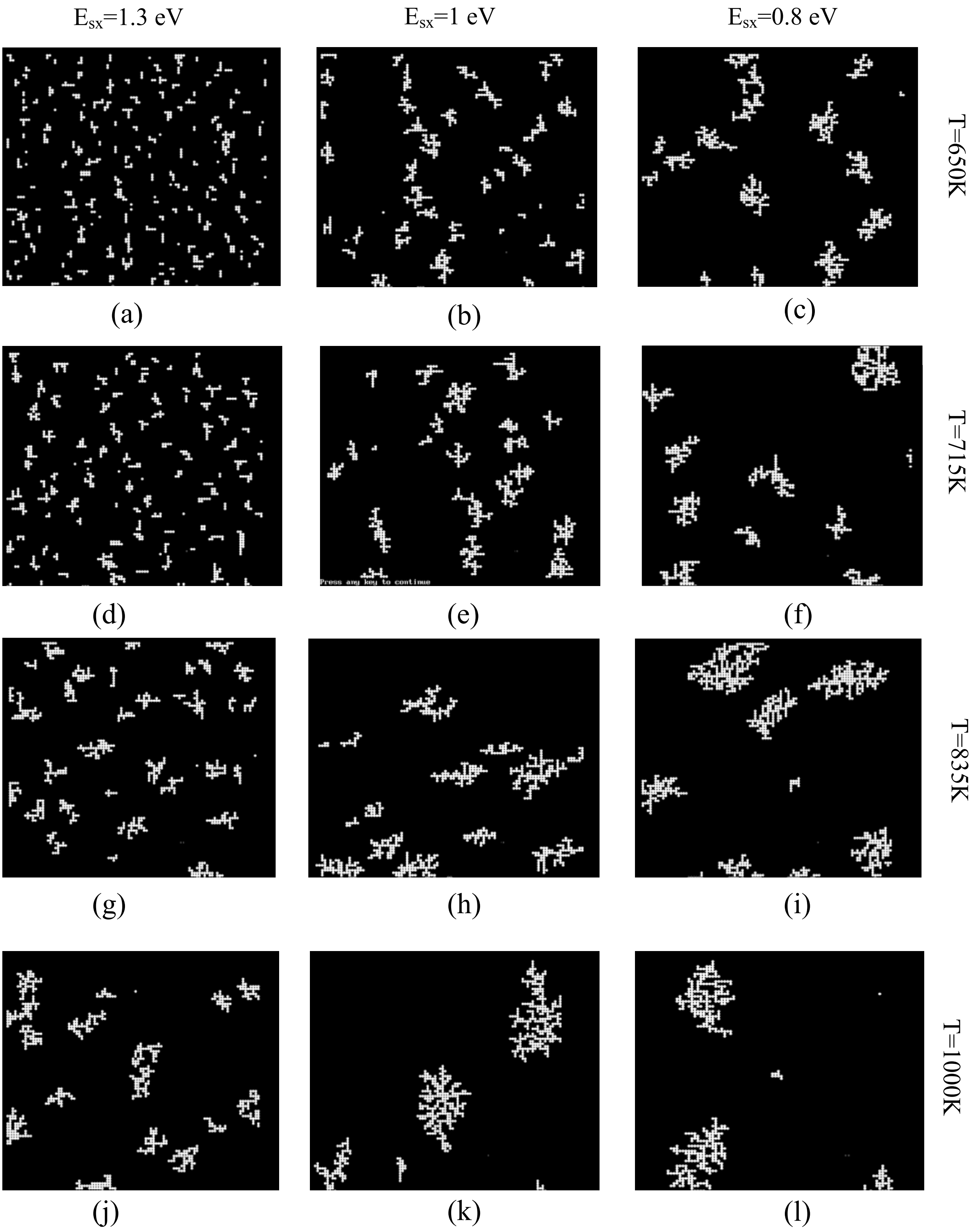

Fig. 1: Section 140*140 of the typical island morphology versus temperature for different values of substrate diffusion energy for model – I with $E_{sy}$ = 1.3 eV, flux F = 0.1 ML/s and Coverage θ = 0.075ML.

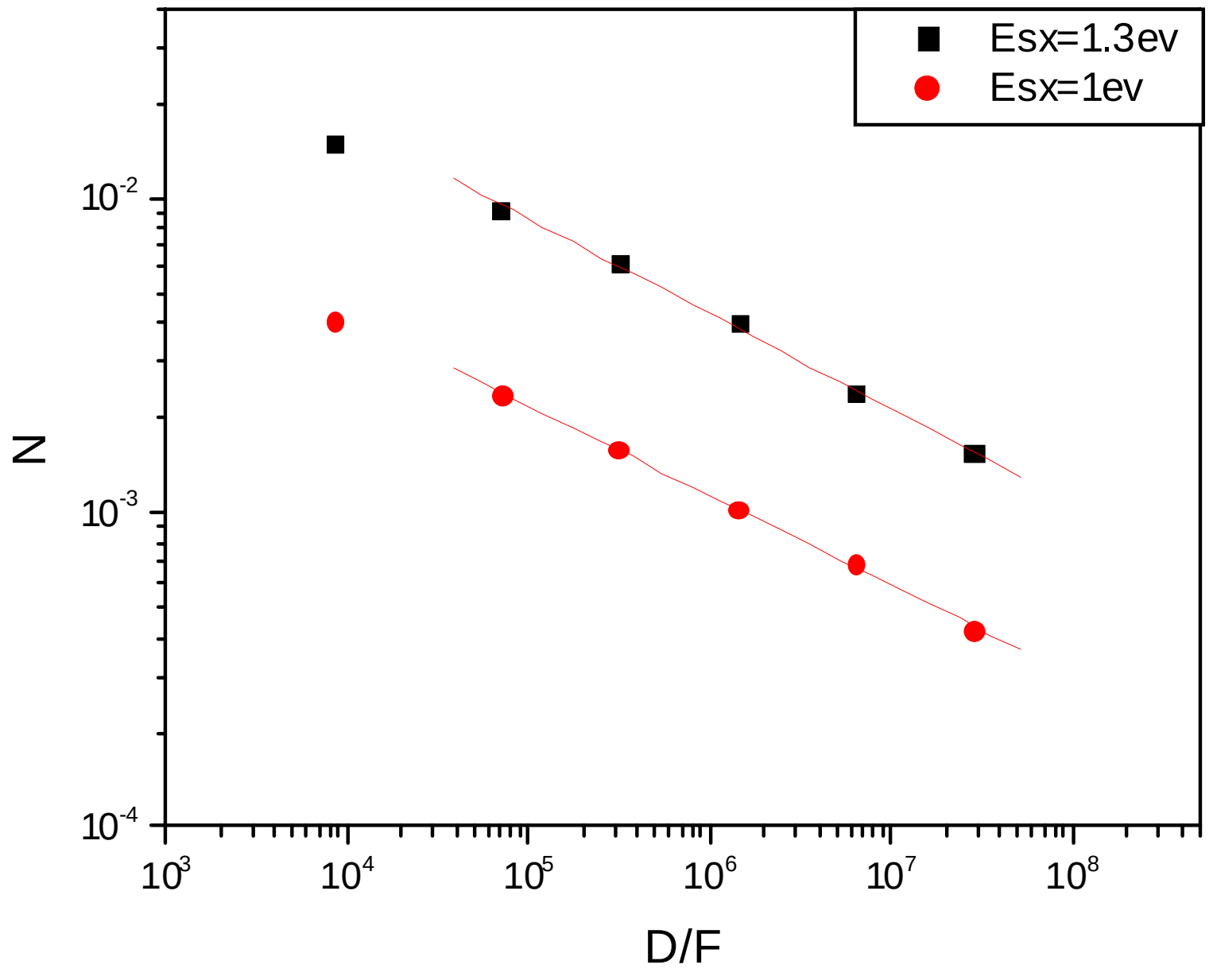

Fig. 2: Island number density as a function of D/F ratio for constant flux F = 0.1 ML/s, Constant Cov = 0.075 ML and Constant $E_{sy}$= 1.3 eV.

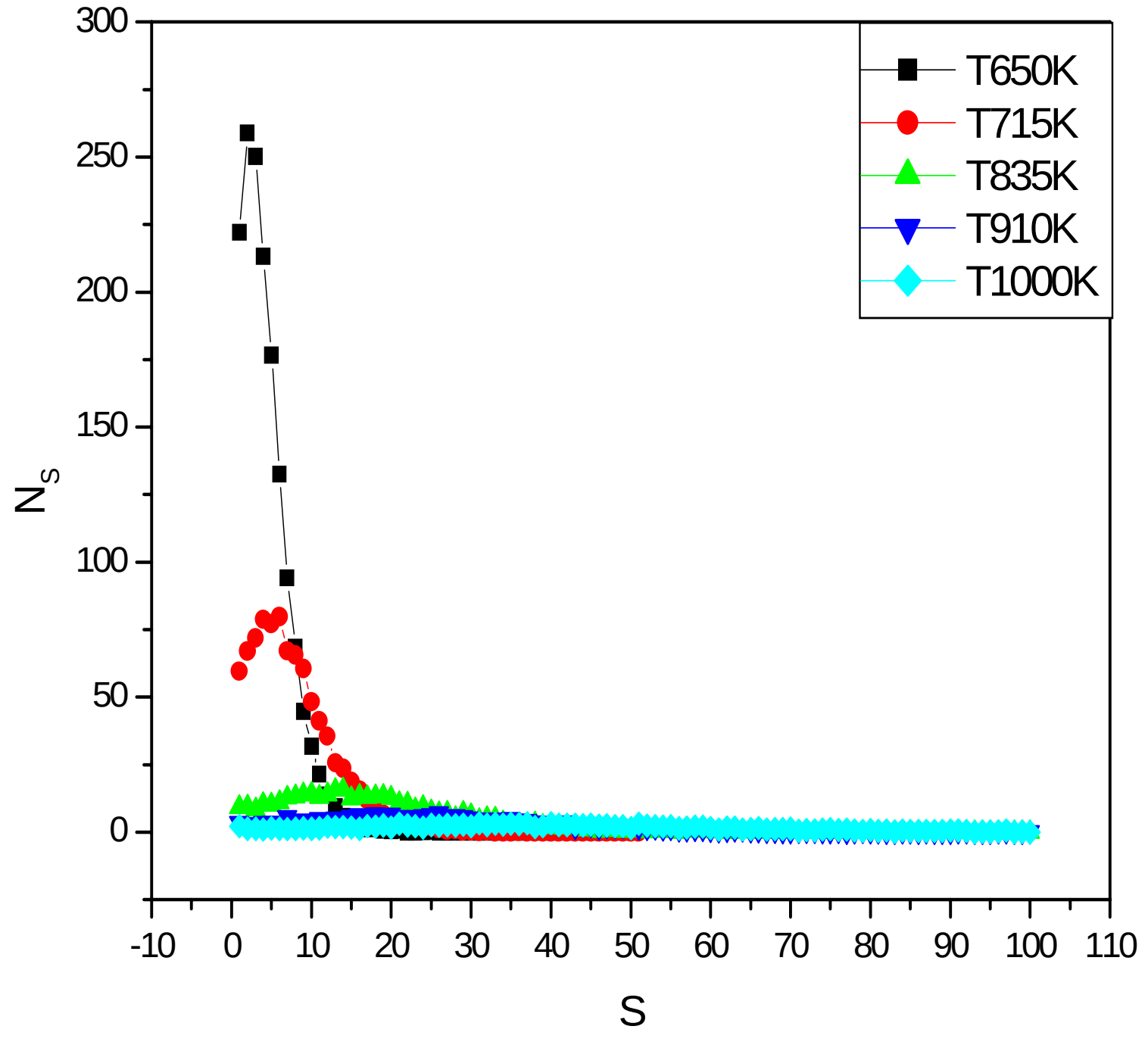

Fig. 3: Island size distribution ($N_s$) versus island size (S) as a function of substrate temperature for **istropic substraste** $E_{sx}$=$E_{sy}$= 1.3 eV, F = 0.1 ML/s, Cov = 0.075 ML.

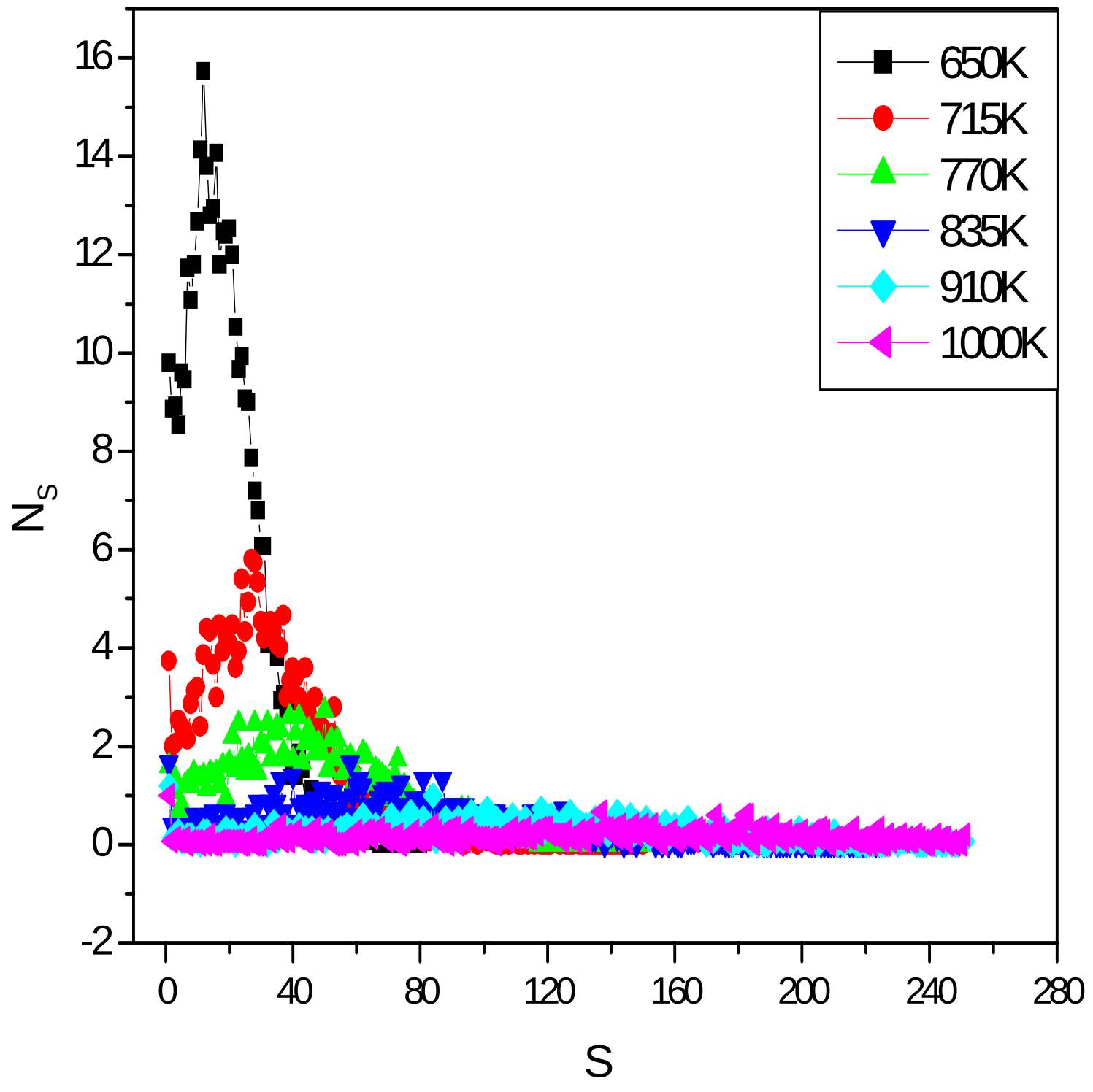

Fig. 4: Island size distribution ($N_s$) versus island size (S) as a function of substrate temperature for **anisotropic substrate** $E_{sx}$= 1 eV, $E_{sy}$= 1.3 eV, F = 0.1 ML/s, Cov = 0.075.

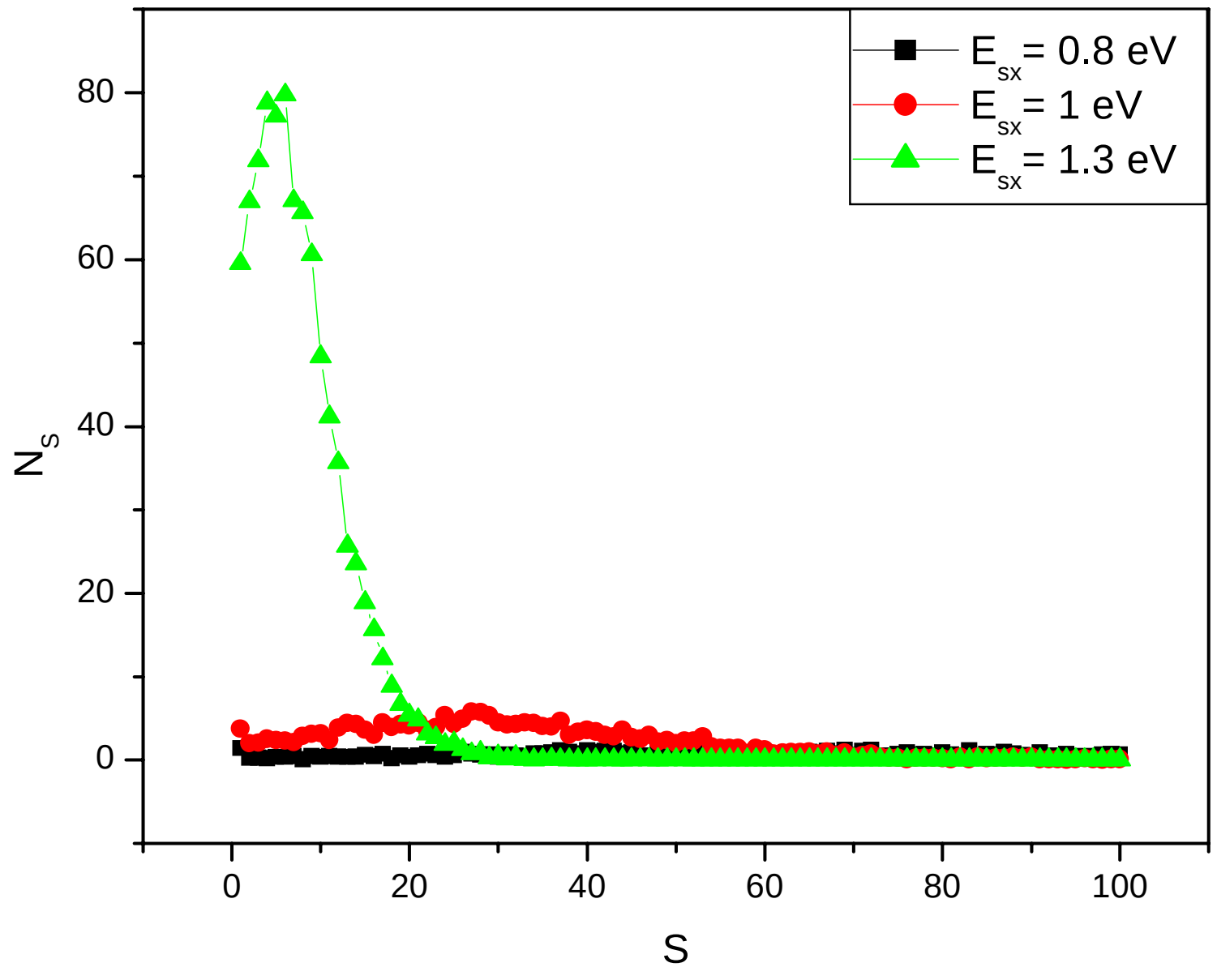

Fig. 5: Comparison of island size distribution ($N_s$) versus island size (S) on different substrates with different in-channel diffusion rate at temperature 715K temperature for F = 0.1 ML/s and Coverage = 0.075 ML.

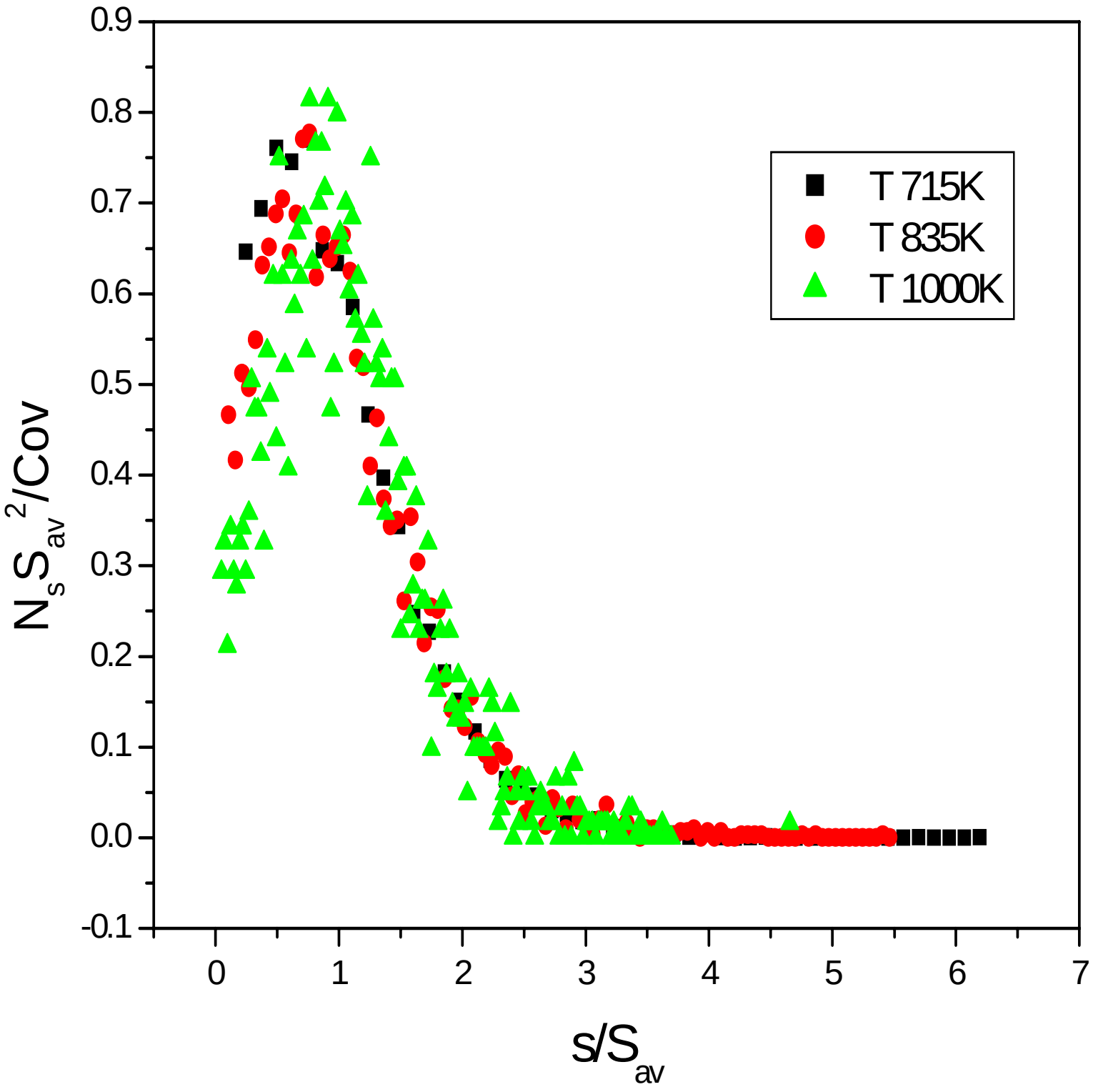

Fig. 6: Plot of scaling function as a function $f(s/S_{av})$ of scaled size $(s/S_{av})$ for Esx = Esy = 1.3 eV at fixed Flux 0.1ML/s, Cov 0.075ML for various temperatures.

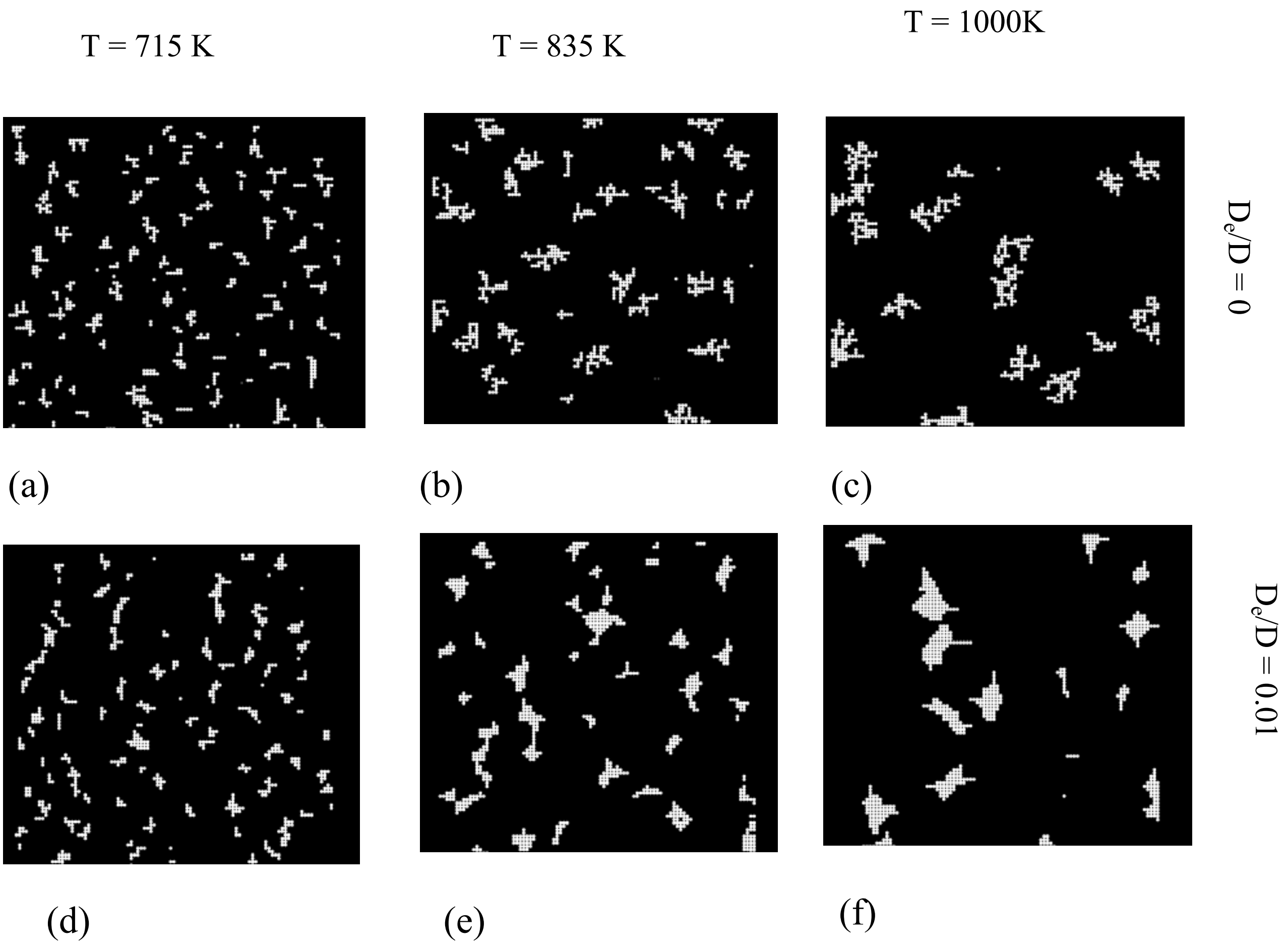

Fig. 7: Section 140*140 of the typical island morphology versus temperature for two different values of $D_e/D$ ratio with $E_{sx}$ = $E_{sy}$=1.3 eV, flux F = 0.1 ML/s and Coverage θ = 0.075ML.

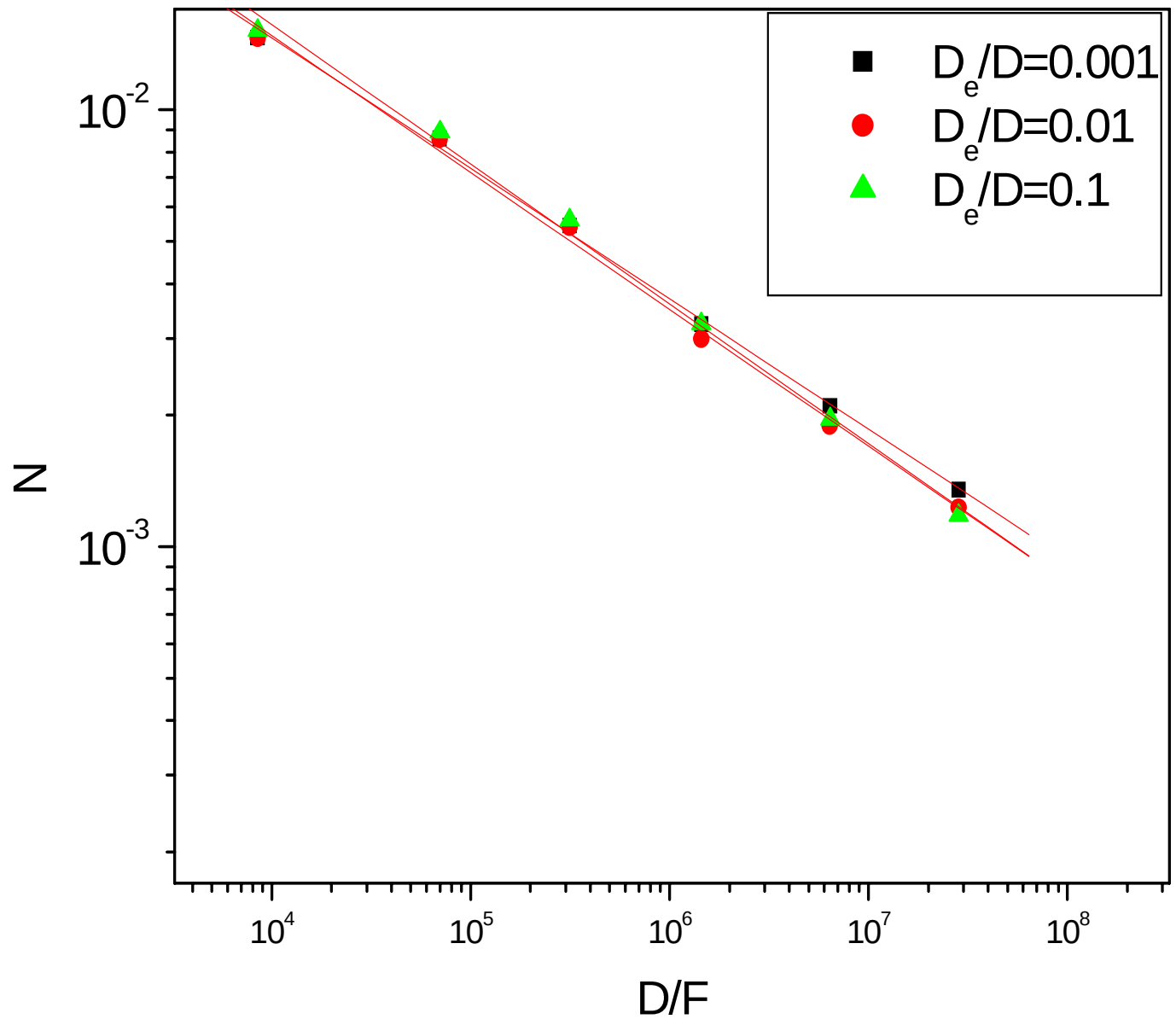

Fig. 8: Variation of island density with D/F for different $D_e/D$ ratio.